\documentclass[12pt]{iopart}
\usepackage{graphicx} 
\usepackage{gensymb}
\usepackage{sidecap}
\usepackage[english]{babel}
\usepackage[utf8]{inputenc}
\usepackage{soul}
\usepackage{xcolor}

\newcommand\hnt[1]{#1}

\begin{document}

\title[QPT of an MS gate via a global beam]{Quantum process tomography of a Mølmer-Sørensen gate via a global beam}

\author{Holly N Tinkey$^1$, Adam M Meier$^1$, Craig R Clark$^1$, Christopher M Seck $^2$, and Kenton R Brown$^1$}

\address{1 Georgia Tech Research Institute, Atlanta, GA, United States of America}
\address{2 Oak Ridge National Laboratory, Oak Ridge, TN, United States of America}
\ead{holly.tinkey@gtri.gatech.edu}
\vspace{10pt}
\begin{indented}
\item[]January 2021
\end{indented}

\begin{abstract}
We present a framework for quantum process tomography of two-ion interactions that leverages modulations of the trapping potential and composite pulses from a global laser beam to achieve individual-ion addressing. Tomographic analysis of identity and delay processes reveals dominant error contributions from laser decoherence and slow qubit frequency drift during the tomography experiment. We use this framework on two co-trapped $^{40}$Ca$^+$ ions to analyze both an optimized and an overpowered Mølmer-Sørensen gate and to compare the results of this analysis to a less informative Bell-state tomography measurement and to predictions based on a simplified noise model. These results show that the technique is effective for the characterization of two-ion quantum processes and for the extraction of meaningful information about the errors present in the system. The experimental convenience of this method will allow for more widespread use of process tomography for characterizing entangling gates in trapped-ion systems.
\end{abstract}

%
\vspace{2pc}

%
%
%
%

\section{Introduction}

\normalsize
Substantial quantum computations in a system of real utility will involve long sequences of one- and two-qubit gate operations \cite{barenco_elementary_1995}, so a thorough characterization of specific gates is important for predicting how they will behave in such sequences and how experimental errors will propagate \cite{knill_quantum_2005}. The most widely used method for quantifying two-qubit entangling gate performance in systems of trapped ions is to use the gate to create a Bell state and then to analyze this state with quantum state tomography (``Bell-state tomography"). This technique is popular, because it requires only the gate interaction and a global one-qubit gate for analysis \cite{benhelm_towards_2008, ballance_high-fidelity_2016}. Bell-state tomography measures how well the experimentally produced state matches an intended entangled state, consolidating coherent and stochastic errors into a single number \cite{bruzewicz_trapped-ion_2019}. However, it falls short of providing a comprehensive description of the quantum process useful for identifying sources of error or for modeling gate performance in the context of more complex sequences.

\hnt{Quantum process tomography (QPT) is a characterization technique that produces a rich representation of a quantum process as a linear mapping of density matrices. This representation can be written as a process matrix that decomposes the dynamics of the experimentally implemented quantum process into a weighted sum over a basis of operators. The process matrix fully captures unitary dynamics, decoherence, and systematic errors for a repeatable quantum process, subject to assumptions about the memory of the environment \mbox{\cite{chuang_prescription_1997, poyatos_complete_1997}}. QPT has been used to characterize one-qubit and two-qubit quantum gates in a variety of physical qubit platforms \mbox{\cite{obrien_quantum_2004, childs_realization_2001,neeley_process_2008, kiesel_linear_2005, riebe_process_2006, herold_universal_2016, navon_quantum_2014, hanneke_realization_2010}}, but the cumbersome experimental demands and time costs have limited its widespread use as a diagnostic tool in trapped-ion systems. Gathering the necessary data in a reasonable timeframe and with enough precision to extract meaningful information about gate errors requires fast one-qubit rotations with low error in comparison to the gate of interest.}

In this paper, we present a framework for performing QPT on two co-trapped ions that alleviates the time, ion-transport, and hardware constraints of other techniques. We achieve single-ion addressing through a method involving modulations of the ions' confining potential, as previously demonstrated by Seck \textit{et al.} \cite{seck_single-ion_2020}, but improved here to reduce the impact on ion temperatures. Section~\ref{sec:QPT} provides a brief summary of the mathematical formulation of QPT and of the challenges of performing QPT with trapped-ion chains. Section~\ref{sec:experiment} contains details about the experimental apparatus, the potential-modulation ion-addressing method, and our QPT framework. \hnt{In Section~{\ref{sec:results}}, we discuss process tomography of identity and delay processes, simulate the effects of various experimental errors in the QPT experiment, and construct the process matrices of both an optimized and an overpowered M\o lmer-S\o rensen (MS) gate \mbox{\cite{sorensen_quantum_1999}}}. The Appendix describes our reduction of heating induced by potential modulations through a finer sampling of the electrode potential waveforms.

\section{QPT with Trapped Ions}
\label{sec:QPT}

QPT is a method for estimating the quantum transfer function of a process that evolves a quantum mechanical state \cite{chuang_prescription_1997}. \hnt{Any quantum process can be expressed as a completely positive linear map $\mathcal{E}$ that acts upon a given input density matrix $\rho_\mathrm{in}$ to produce the output state $\rho_\mathrm{out}$: $\mathcal{E}(\rho_\mathrm{in}) = \rho_\mathrm{out}$. Mathematically, the map $\mathcal{E}$ can be expressed using an operator sum representation over a set of Kraus operators $\hat{A}_{m,n}$ with complex weights stored in the process matrix denoted with Greek letter $\chi$: $\mathcal{E}(\rho_\mathrm{0}) = \Sigma_{mn}\chi_{mn}\hat{A}_m\rho_\mathrm{0}\hat{A}^\dagger_n$ \mbox{\cite{dariano_quantum_2001}}. For a given set of Kraus operators, typically tensor products of I, X, Y, and Z Pauli operators for the qubits, a quantum process can be fully described by the elements of the process matrix \mbox{\cite{kofman_two-qubit_2009}}.} \hnt{For a repeatable process acting on $n$ qubits, and subject to assumptions about the qubits' environment, $4^n$ $\times$ $4^n$ elements are sufficient for $\chi$ to provide a complete mathematical picture of the quantum process under study \mbox{\cite{wolk_distinguishing_2019}}.}

QPT experiments involve initializing $n$ qubits into a set of $4^n$ input states, acting on these states with the quantum process, and performing state tomography on the resulting output states. State tomography of each state can be performed using \hnt{a set of $4^n$ different measurement settings}, each of which can be implemented by performing a different non-entangling gate on the qubits prior to standard (Z-basis) measurement of the qubits. In this work, we use maximum likelihood estimation (MLE) to derive the most probable process matrix given our experimentally measured populations, thereby avoiding unphysical process matrices that can be obtained through the simpler technique of matrix inversion when noise is present in the measurements \cite{jezek_quantum_2003, blume-kohout_optimal_2010, van_enk_when_2013, Hradil_MLE_2004}. After constructing $\chi$, it becomes possible to compare the experimental process to the ideal quantum gate, to calculate the process fidelity, and to identify errors in the process.

To date, experimental \hnt{demonstrations of process tomography} on entangling gates for trapped ions have leveraged several different techniques to implement the required individual addressing of the ions; Riebe \textit{et al.} characterized cNOT gate sequences with and without pulse shaping using tightly focused beams to produce single-ion rotations, a technique which produced crosstalk effects on the neighboring ions \cite{riebe_process_2006}. \hnt{Herold \mbox{\textit{et al.}} also studied a cNOT gate sequence using narrowly focused beams to provide cascading pairwise addressing for single-ion control, but the cascaded transport time and crosstalk effects produced 5\% error in a two-ion identity process studied with QPT mbox{\cite{herold_universal_2016}}. Navon \textit{et al.} investigated the performance of concatenated MS gates to extract an error per gate by using a micromotion-induced transition for individual addressing; this method is susceptible to drift of the background electric field that contributed to an error of 5\% for an identity process \mbox{\cite{navon_quantum_2014}}}. A study by Hanneke \textit{et al.} relied on separating ions from larger chains and on sympathetic cooling with a second ion species to perform individual rotations, a technique which is time intensive and requires additional hardware to support the second ion species \cite{hanneke_realization_2010}. 

\hnt{QPT analysis in its simplest form assumes perfect initialization of the input states and perfect measurement of the outputs. Because different initializations and measurements are typically implemented with single-ion rotations on each ion before and after the gate under test, the difficulty in performing fast, high-fidelity individual-ion addressing has been a barrier to more widespread adoption of QPT in trapped-ion systems. Ideal single-ion rotations should have high fidelity with respect to the gate being characterized, so that analysis of the gate is not obscured by individual addressing errors, and should be fast to limit the effects of environmental decoherence.} The potential modulation method we utilize for single-ion addressing provides this critical combination of fidelity and speed. \hnt{We incorporate the method into an experimental framework to demonstrate QPT analysis of an entangling MS gate and study the behavior of identity and delay processes to identify relevant experimental errors.} Implementing this method requires simple hardware to control the trap electrode voltages and the phase of a single global beam.

\hnt{We partially discriminate the errors of our individual addressing technique from those of the MS gate by using an approach that follows along the same path as gate set tomography (GST \mbox{\cite{blume_GST_2017}}) but does not go as far in rigor or in complexity. Our results indicate that the potential modulation approach would be suitable also for increasing the ease and speed of GST and related protocols, and the use of these protocols would likely allow an even more complete accounting of our error sources. However, we believe that the relative simplicity of QPT gives it a lasting place in the quantum characterization toolbox.}

\section{Experiment}
\label{sec:experiment}

For these experiments, two $^{40}$Ca$^+$ ions are confined with a surface-electrode linear Paul trap \cite{seck_single-ion_2020, shappert_spatially_2013}. Radial confinement is provided by an RF potential at 56.4 MHz (peak voltage $\approx$ 176 V) that traps the ions 58 $\mu$m above the trap surface. An arbitrary waveform generator (AWG) with a 5 ns sampling rate applies synchronized potentials to 42 segmented electrodes along the trap axis and implements waveforms for ion transport and confinement modulation. The center-of-mass (COM) and breathing mode (BM) secular frequencies are $\omega_\mathrm{COM}$ = 2$\pi$ $\times$ 1.41 MHz and $\omega_\mathrm{BM}$ = 2$\pi$ $\times$ 2.45 MHz respectively. Qubit information is stored in the $|S\rangle \equiv |S_{1/2}, m_j = -1/2\rangle$ ground state and metastable $|D\rangle \equiv |D_{5/2}, m_j = -1/2\rangle$ state, and transitions between these states are achieved with a narrow-linewidth 729 nm beam oriented at 45$\degree$ to the trap axis and illuminating both ions. The ion states are measured by illuminating them with light near 397 nm and monitoring the resulting fluorescence; we can distinguish between the two-ion bright state ($|SS\rangle$), the one-ion bright subspace (\hnt{superpositions} of $|SD\rangle$ and $|DS\rangle$), and the dark state ($|DD\rangle$). Repeating an experiment allows us to \hnt{determine} the fraction of the final population ($P_2$, $P_1$, $P_0$) in each of these three possible configurations.

We realize individually addressed one-qubit rotations through composite sequences of 729 nm laser pulses and modulations to the confining potential as described in \cite{seck_single-ion_2020}. \hnt{Here we use two laser pulses and two potential modulations to produce one-qubit rotations on a single ion in 25 $\mu s$: (1) a laser pulse produces half of the intended qubit rotation, (2) a change in potential confinement creates a differential optical phase shift of $\pi$ between the ions, (3) a second laser pulse completes the intended rotation on the targeted ion while returning the untargeted ion to its initial state, and (4) the confinement is returned to its initial value for the next operation.} For our beam geometry and orientation, \hnt{a multiplicative scaling $\mathcal{S}$ of trap electrode voltages produces} center-of-mass frequency $\omega_\mathrm{COM}^\prime$ = 2$\pi$ $\times$ 1.71 MHz and reduces the ion separation by $0.531$ nm, creating a differential optical phase shift between the ions of $\pi$ in the global 729 nm beam. Our previous work with these sequences was limited by significant heating during the potential modulations; here the incorporation of AWGs with a much higher sampling rate allows for faster and smoother potential modulations that reduce this heating \hnt{(further details in the Appendix)}.

\begin{figure}
    \centering
    \includegraphics[width=0.8\textwidth]{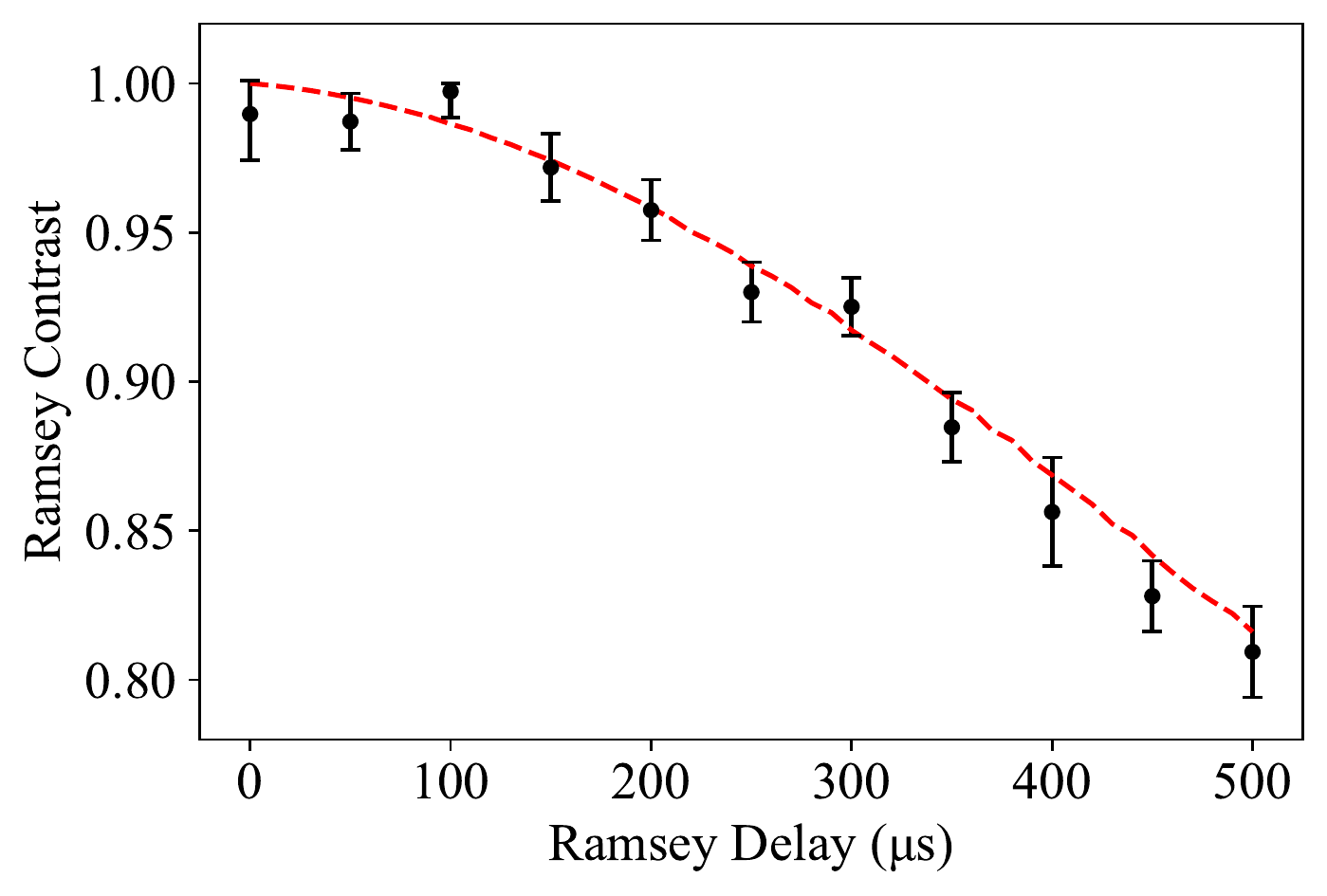}
    \caption {\label{fig:ramsey}\hnt{Ramsey contrast as a function of delay between pulses: black points represent experimentally measured values. The red dashed line shows a laser noise model fit to the Ramsey measurements. Error bars represent uncertainty in the fit used to determine Ramsey contrast for a given delay.}}
\end{figure}

The MS gate under analysis in this work is performed by the simultaneous application of two 729 nm tones at detunings $\pm 2\pi\times 8.3$~kHz from the red and blue breathing-mode motional sidebands of the qubit transition for a duration of 120 $\mu$s. \hnt{To compensate for Stark shift effects due to the intensity of the gate beams, the red and blue sideband frequencies are calibrated with the other tone illuminating the ion at a detuning of 30~kHz}. The optical intensity is adjusted to produce a maximally entangled state $(|SS\rangle - i|DD\rangle ) / \sqrt{2}$. We estimate the Bell-state fidelity of the gate using the results of two experiments:  a measurement of the populations $P_0$ and $P_2$ after the gate is performed, and a measurement of the populations as a function of the phase of a final analysis pulse to construct a parity signal with amplitude $P_\mathrm{amp}$ \cite{leibfried_experimental_2003, akerman_universal_2015}. The Bell state fidelity for our optimized gate is $F_\mathrm{BST} = \frac{1}{2}P_\mathrm{amp} + \frac{1}{2}(P_0 + P_2) = $ 96.2(7)\%. \hnt{The gate fidelity is limited largely by the $730$ $\mu \mathrm{s}$ coherence time of the laser. A laser noise model fit to our single-ion Ramsey experiments (shown in Fig.~{\ref{fig:ramsey}} and discussed in detail in Section~{\ref{sec:identity}}) predicts a loss of 1.8\% Ramsey contrast for one ion during a $120$ $\mu \mathrm{s}$ delay, translating to 3.6\% error for a two-ion process.} 

Our experimental QPT framework incorporates the potential modulation-based single-ion rotations to realize state initialization before the quantum process under analysis and measurement rotations for state tomography after the quantum process. One of four rotations (identity, $\pi$ and $\frac{\pi}{2}$ rotations about the X axis, and $\frac{\pi}{2}$ about the Y axis) is applied to each ion after initialization in the $|SS\rangle$ state, creating one of 16 input states ($\rho_\mathrm{in}$) on which the quantum process will act. To characterize the state ($\rho_\mathrm{out}$) after the process, one of the same four rotations is applied to each ion to generate one of 16 readout directions. In total we perform 256 different experimental sequences, one for each combination of initialization and measurement settings, and we repeat each of these experiments 500 times to develop statistics. \hnt{Fitting our two-ion fluorescence histograms to a sum of Poissonians allows us to determine the populations $P_2$ for each sequence while avoiding the lengthy transport operations that would be required for individual-ion detection with a single PMT. However, this detection strategy requires $4^n$ $\times$ $4^n$ measurements to extract the full process matrix.}

\section{QPT Results and Analysis} 
\label{sec:results}

\hnt{For all of the following analyses we determine the quantum process $\chi_\mathrm{exp}$ most likely to have generated our experimental data using an MLE algorithm closely following the outline of Hradil \mbox{\textit{et al.}} \mbox{\cite{Hradil_MLE_2004}}. No corrections for state preparation or measurement errors have been applied to the data.} We calculate the process fidelity $F_\mathrm{p} = \mathrm{Tr}[\chi_\mathrm{exp} \cdot \chi_\mathrm{ideal}]$ through comparison with the expected process $\chi_\mathrm{ideal}$ \cite{riebe_process_2006}. We estimate the uncertainty on this fidelity through a resampling of our experimental data based on the statistical uncertainty for each population measurement.

\subsection{Identity and Delay Process Tomography}
\label{sec:identity}

\begin{figure}
    \centering
    \includegraphics[width=1.0\textwidth]{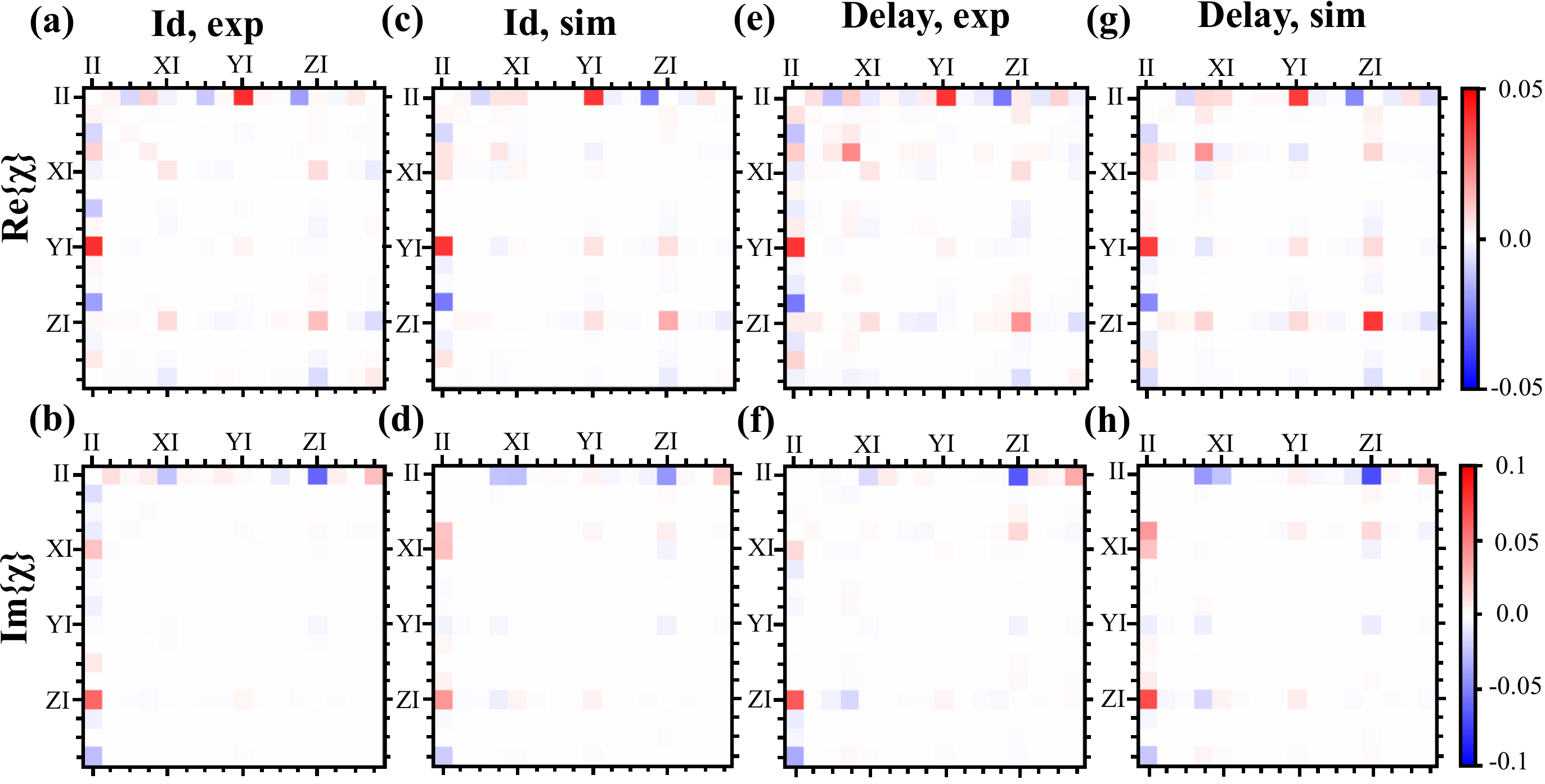}
    \caption {\label{fig:chi_id_delay}\hnt{(a) Real and (b) imaginary amplitudes of the process matrix $\chi_\mathrm{Id}$ constructed from experimental data of an identity operation with process error 3.2(3)\%. Elements of $\chi$ are in the basis $\sigma_i^\mathrm{ion 1} \otimes \sigma_j^\mathrm{ion 2}$ where $\sigma_\mathrm{I,X,Y,Z}$ are denoted with I, X, Y, and Z and are in order II, IX, IY, IZ, XI, XX, ... ZZ. (c) Real and (d) imaginary amplitudes of the process matrix $\chi_\mathrm{Id}$ constructed from a simulated data set for an identity operation with process error 3.2\%. (e) Real and (f) imaginary amplitudes of the process matrix $\chi_\mathrm{Del}$ constructed from experimental data of a 120 $\mu$s delay with error 6.7(4)\%. (g) Real and (h) imaginary amplitudes of the process matrix $\chi_\mathrm{Del}$ constructed from a simulated data set for a 120 $\mu$s delay with error 7.2\%. The $\chi^\mathrm{II-II}$ amplitudes have been removed in the image to show the error amplitudes more clearly.}}
\end{figure}

We first apply the QPT experimental framework to an identity (null) operation and to a delay (time-evolution) operation to understand the errors present in our system and to examine how these errors accumulate during the tomography procedure. The experiment for the identity operation consists of performing the single-ion initialization and measurement rotations with no intermediate interaction or delay. In the absence of decoherence and systematic experimental errors, all of the \hnt{amplitude} for this process matrix $\chi_{\mathrm{Id}}$ [Fig.~\ref{fig:chi_id_delay} (a) and (b)] would fall into  \hnt{$\chi_\mathrm{Id}^\mathrm{II,II}$ [i.e., the Kraus operator element that acts on $\rho_\mathrm{in}$ as ($\mathrm{I}_{1}\mathrm{I}_{2} )\rho_\mathrm{in} (\mathrm{I}_{1} \mathrm{I}_{2})$]}. Here we determine the fidelity is \hnt{$F_\mathrm{p} = $ 96.8(3)\%}. \hnt{Our laser noise model predicts a single-ion Ramsey contrast loss of 1.3\% for the 100~$\mu$s duration of the initialization and tomography pulses, which translates into a two-ion process error of 2.6\% (Fig.~{\ref{fig:ramsey}}). Our measured process error [3.2(3)\%] is slightly lower than the identity process error of 5\% measured in previous studies employing different individual addressing methods \mbox{\cite{herold_universal_2016, navon_quantum_2014}}.}

The experiment for the delay operation consists of the same initialization and measurement rotations now separated by a 120 $\mu$s delay to mimic the duration of our MS gate. QPT analysis here produces a process matrix [$\chi_\mathrm{Del}$, Fig.~\ref{fig:chi_id_delay} (e) and (f)] with \hnt{process fidelity 93.3(4)\%}. \hnt{This process error magnitude agrees with a naive combination of the 3.2(3)\% error measured in the identity process and the 3.6\% error predicted for a 120 $\mu$s delay by our laser noise model (see Fig.~{\ref{fig:ramsey}} and Section~{\ref{sec:experiment}}).}

\hnt{These identity and delay process matrices reveal prominent errors in similar elements. There are large, real amplitudes ($\chi^\mathrm{II,IY}$, $\chi^\mathrm{II,YI}$, and $\chi^\mathrm{II,YZ}$) along the II column and row in Y-type process elements as well as diagonal Z-type process matrix elements ($\chi^\mathrm{IZ,IZ}$ and $\chi^\mathrm{ZI,ZI}$) that grow in amplitude with delay. The imaginary amplitudes of the Z-type process matrix elements ($\chi^\mathrm{II,IZ}$ and $\chi^\mathrm{II,ZI}$) also increase between identity and delay processes.}

\hnt{These non-zero amplitudes represent errors accrued during state preparation and state tomography that are mistakenly attributed to $\chi$ elements during mathematical reconstruction. The QPT formulation assumes that the expected input density matrix $\rho_\mathrm{in}$ has been created perfectly and that the results density matrix $\rho_\mathrm{out}$ has been measured perfectly. Errors accumulated during individual addressing do not all commute with the qubit rotations being performed, so the effects of various errors on the process matrix are not straightforward and depend heavily upon experimental details of the individual addressing method. Simulating the identity and delay processes with the inclusion of routine errors such as laser phase noise and qubit frequency drift \mbox{\cite{riebe_process_2006, wang_demonstration_2010}} as well as errors more specific to our system provides greater insight into how these errors manifest in the process matrices.}

\hnt{We simulate data sets for our QPT identity and delay experiments using rotation operators $\hat{R}(\theta,\phi) = \mathrm{exp} [\frac{-i\theta}{2}(\sigma^x \mathrm{cos}\phi +  \sigma^y \mathrm{sin}\phi)]$ to describe the action of all 729 nm laser pulses used in the individual addressing composite sequences for state initialization and tomography. Errors in pulse duration are realized as deviations from the ideal $\theta$ for each pulse, while errors in the phase are applied as deviations from the ideal $\phi$ of each pulse and are propagated to subsequent pulses. We simulate each of the 256 unique initialization and tomography sequences 500 times, with $\theta$ and $\phi$ chosen from appropriate random distributions (described below), to replicate the QPT experiment as closely as possible. Here we allow for errors that occur on different time scales during the course of an experiment. Constant errors in $\theta$ and $\phi$ represent miscalibrations of the individual addressing parameters and are applied to all relevant pulses. Qubit and laser frequency errors are implemented by changing $\phi$ linearly depending on the timing of a given pulse within a sequence. In this case, slow frequency drift (7.0 Hz/min) is modeled as a frequency offset which increases (or decreases) linearly in time but which remains constant during all 500 repetitions of a sequence. Faster frequency variations are simulated by pulling from a fixed-width Gaussian distribution (300 Hz standard deviation) once for each iteration of a given sequence. Similarly, fast phase noise is implemented by pulling from a different Gaussian distribution whose standard deviation  depends on the delay between pulses (0.015 radians of standard deviation per $\sqrt{\mu\mathrm{s}}$ delay). Slow frequency drift was estimated from measurements of the qubit frequency before and after the QPT identity and delay experimental runs. The fast frequency variation and phase noise parameters were extracted from a fit of this laser noise model to independent Ramsey decoherence measurements (see Fig.~{\ref{fig:ramsey}}). The predicted population $P_2$ for each sequence is averaged over all 500 iterations to create a simulated data set, and from this we calculate a process matrix with the MLE algorithm.}

\hnt{These simulations allow us to predict the impact of laser noise on our QPT experiments. The fast phase variations, fast phase noise, and slow qubit frequency drift described above predict amplitudes of Z-type elements  that agree well with the amplitudes seen in experimental identity and delay $\chi$ matrices (Fig.~{\ref{fig:chi_id_delay}}). In particular, these sources of error recreate the amplitudes of $\chi^\mathrm{IZ,IZ}$, $\chi^\mathrm{ZI,ZI}$, $\chi^\mathrm{ZI,II}$, and $\chi^\mathrm{II,ZI}$ that grow in magnitude from identity to delay process. Slow qubit frequency drift causes larger Z-type amplitudes for ion 2 compared to ion 1 (e.g. $\mathrm{Im}[\chi^\mathrm{II,ZI}] > \mathrm{Im}[\chi^\mathrm{II,IZ}]$). This can be understood as an artifact of a time-dependent error growing during an experimental data run with a fixed order of sequences. Because our QPT experiment uses a fixed progression of initialization and tomography sequences, frequency drift will impact later sequences more than early ones.  Simulations using laser noise and frequency drift alone can explain the growth in overall process error due to the delay, but they do not explain the full process error magnitude or the large $\chi^\mathrm{II,YI}$, $\chi^\mathrm{YI,II}$, $\chi^\mathrm{II,YZ}$, and $\chi^\mathrm{YZ,II}$ amplitudes observed in experiment [Fig.~{\ref{fig:chi_id_delay}} (a) and (e)].}

\hnt{Of the individual addressing error sources we considered incorporating into the simulations, we find that only two create large Y-type error amplitudes that impact the ions differently. The potential modulation method depends both on a multiplicative scaling $\mathcal{S}$ of the trap electrode voltages, which is calibrated to adjust the inter-ion distance, and on a position phase $\phi_p$, which corrects for the shift in position of ion 1 with respect to the 729 nm beam wavefronts when the potential is scaled. When $\mathcal{S}$ is miscalibrated, ion 2 does not receive the expected differential phase shift of $\pi$ in the scaled potential, and simulations predict prominent amplitudes in Y-type elements for ion 2. When $\phi_p$ is incorrect, an inaccurate position phase is used for every laser pulse in the scaled potential, and simulations predict prominent amplitudes in Y-type elements for both ions. Combinations of these miscalibrations create imbalanced IY, YI, YZ, and ZY errors that do not depend on the delay between initialization and tomography pulses. Simulations using a $\phi_p$ error of -145~mrad along with a fractional $\mathcal{S}$ error of 1.9\% (creating a +155~mrad phase error for ion 2) reproduce the observed $\chi^\mathrm{II,YI}$ and $\chi^\mathrm{II,YZ}$ amplitudes [Fig.~{\ref{fig:chi_id_delay}} (a),(c) and (e),(g)]. These individual addressing errors create very specific amplitude signatures and can be mitigated with more careful calibration.}

\subsection{MS Gate Process Tomography}

\begin{figure}
    \centering
    \includegraphics[width=0.7\textwidth]{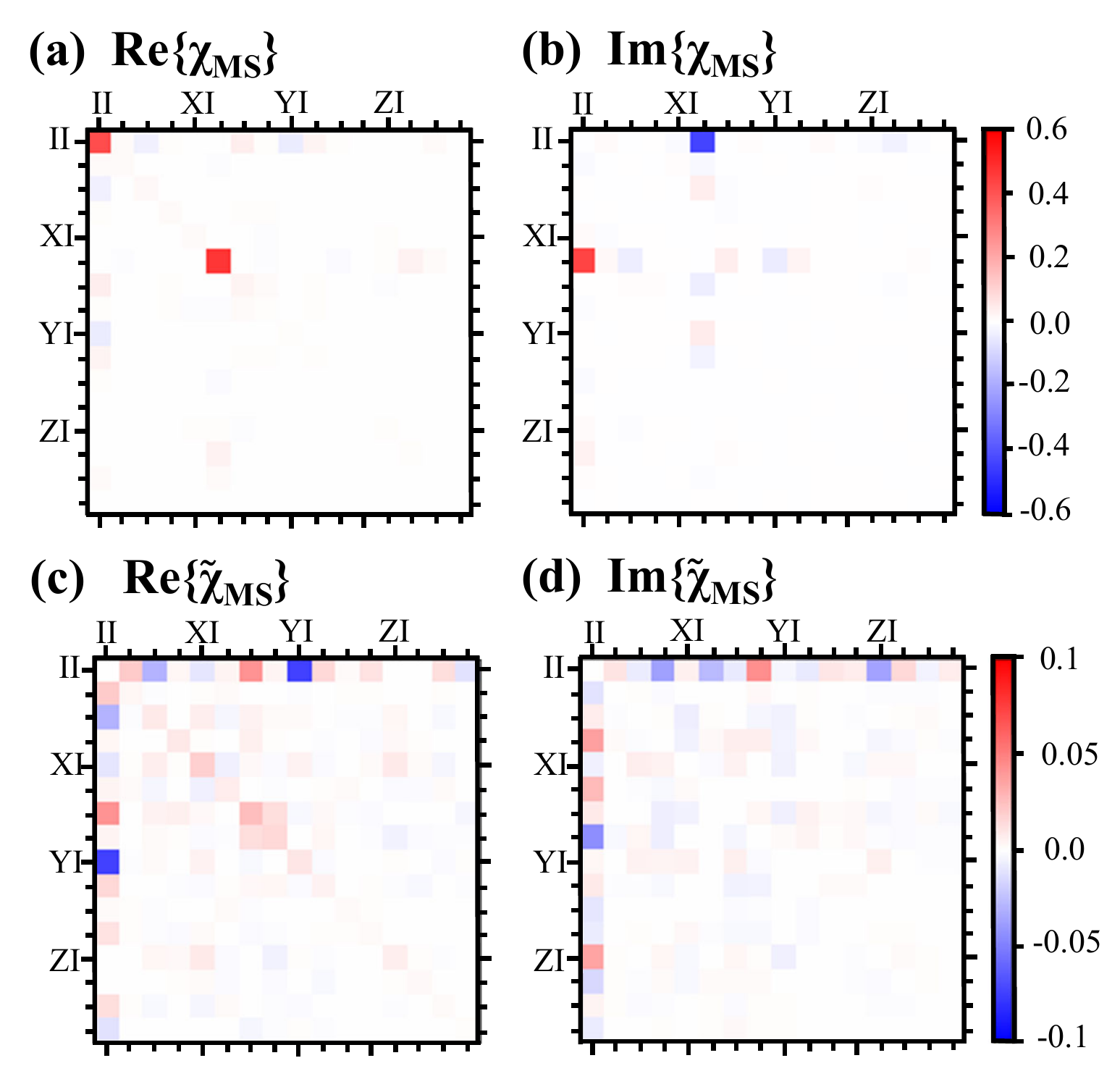}
    \caption {\label{fig:chi_MS}(a) Real and (b) imaginary components of the process matrix $\chi_\mathrm{MS}$ of an experimentally implemented MS gate with $F_\mathrm{p} =$ 88.1(5)\%. (c) Real and (d) imaginary components of the error process matrix $\tilde{\chi}_\mathrm{MS}$ for the MS gate. \hnt{The $\tilde{\chi}_\mathrm{MS}^\mathrm{II-II}$ amplitude has been removed in the image to show the error amplitudes more clearly.}}
\end{figure}

\hnt{Before performing QPT of the MS gates, we calibrate both the phase of the applied MS pulse and the phase of the following tomography pulses to account for Stark shifts during the gate.} The propagator for the MS gate can be written as $U_\mathrm{MS}=\mathrm{exp}(-i\frac{\pi}{4} \mathrm{X}_1 \mathrm{X}_2) = (\mathrm{I}_1 \mathrm{I}_2 -i \mathrm{X}_1 \mathrm{X}_2)/\sqrt{2}$, so that the corresponding $\chi$ process matrix contains four nonzero elements: \hnt{$\chi_\mathrm{MS}^\mathrm{II,II} = \chi_\mathrm{MS}^\mathrm{XX,XX} = \frac{1}{2}$ and $\chi_\mathrm{MS}^\mathrm{XX,II} = -\chi_\mathrm{MS}^\mathrm{II,XX} = \frac{i}{2}$ \mbox{\cite{navon_quantum_2014}}}. The experimentally determined process matrix for our MS gate [$\chi_\mathrm{MS}$, Fig.~\ref{fig:chi_MS}(a) and (b)] closely agrees with these predicted values, although there is some disagreement due to \hnt{laser noise and individual addressing errors (discussed in Section~{\ref{sec:identity}}) }and due to errors inherent in the gate itself. Here we calculate a process fidelity $F_\mathrm{p} = $ 88.1(5)\%. \hnt{This 11.9(5)\% deviation from the ideal gate can be understood roughly as a combination of the 6.7(4)\% error measured for the delay operation, capturing errors in QPT, and the 3.8(7)\% MS gate error determined through Bell-state tomography, capturing errors in the gate (Section~{\ref{sec:experiment}}). However, this addition of errors is not a rigorous analysis: while Bell-state infidelity captures some of the errors of the MS gate, it does not account for others, such as systematic errors in the gate phase calibration. Moreover, adding the delay infidelity to the Bell-state infidelity likely double counts errors due to laser decoherence.}

\hnt{To clarify the error analysis, the MS results may be recast by writing the measured quantum process as a composition, $\chi_\mathrm{MS}= \chi_\mathrm{MS, ideal} \circ \tilde{\chi}_\mathrm{MS}$, of the ideal unitary evolution $\chi_\mathrm{MS, ideal}$ and an error process $\tilde{\chi}_\mathrm{MS}$ [Fig.~{\ref{fig:chi_MS}} (c) and (d)] \mbox{\cite{dewes_characterization_2012, korotkov_error_2013}}. Recasting the results in terms of $\tilde{\chi}_\mathrm{MS}$ makes it easier to visualize errors that might otherwise be overshadowed by the ideal behavior in the usual process matrix. Most notably, we observe prominent real amplitude in $\tilde{\chi}_\mathrm{MS}^\mathrm{II-YI}$ indicating a possible $\phi_p$ calibration error.}

\begin{figure}
    \centering
    \includegraphics[width=0.7\textwidth]{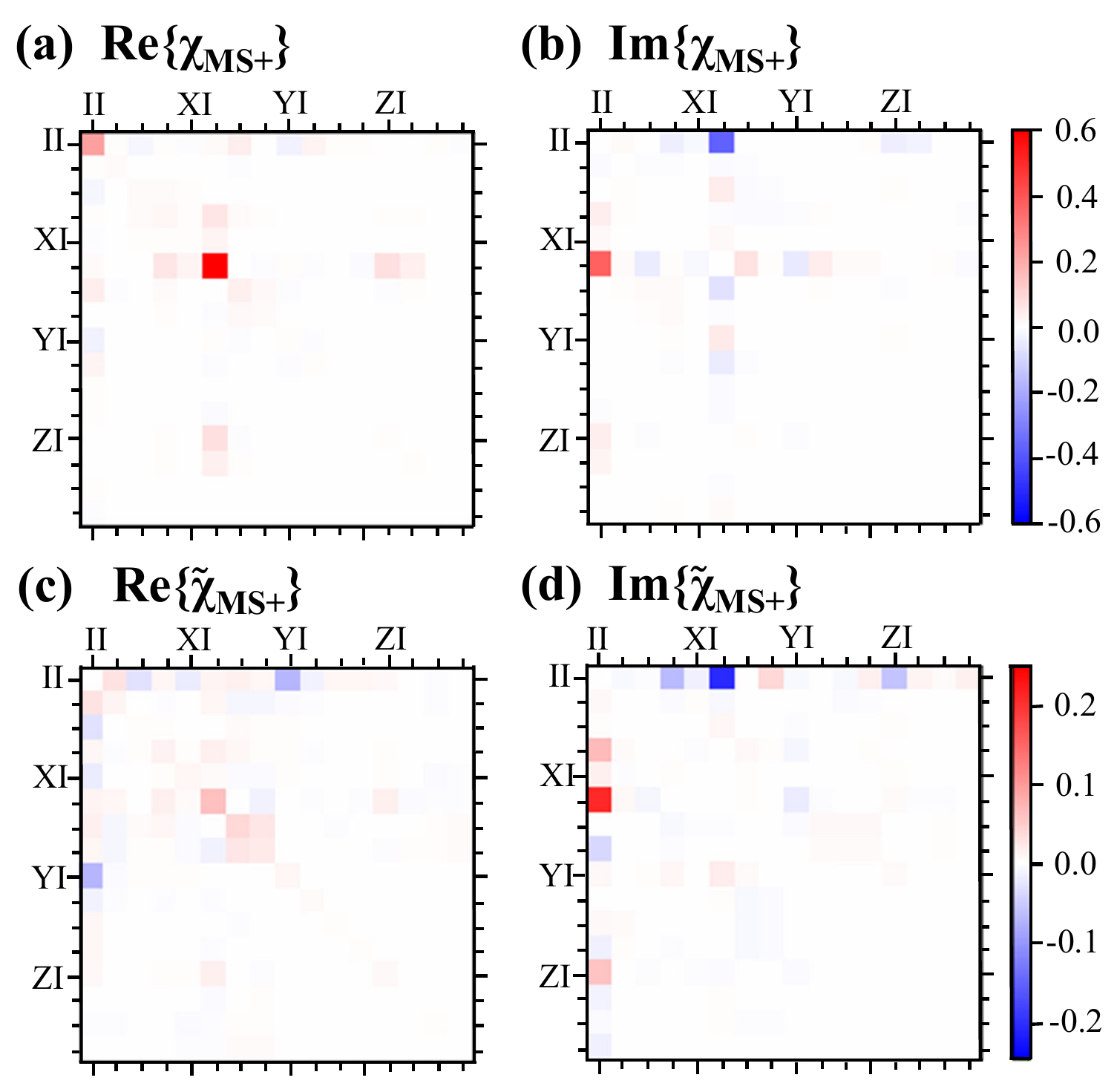}
    \caption {\label{fig:chi_MS+}(a) Real and (b) imaginary components of the process matrix $\chi_\mathrm{MS+}$ for an overpowered MS gate with $F_\mathrm{p} = $81.5(6)\%. (c) Real and (d) imaginary components of the process error matrix $\tilde{\chi}_\mathrm{MS+}$ for the overpowered MS gate. \hnt{The $\chi_\mathrm{MS}^\mathrm{II,II}$ amplitude has been removed in the image to show the error amplitudes more clearly.}}
\end{figure}

\hnt{To emphasize the diagnostic power of QPT, we intentionally increase the RF power driving an acousto-optic modulator in the MS gate beam path by 1.5 dB. Process tomography of the overpowered gate produces a process matrix $\chi_\mathrm{MS+}$ with imbalanced real amplitudes ($\chi_\mathrm{MS+}^\mathrm{II,II} \neq \chi_\mathrm{MS+}^\mathrm{XX,XX}$) and lowered imaginary amplitudes $\chi_\mathrm{MS+}^\mathrm{XX,II}$ and $\chi_\mathrm{MS+}^\mathrm{II,XX}$ [Fig.~{\ref{fig:chi_MS+}}(a) and (b)] . We calculate a process fidelity of 81.5(6)\% with respect to an ideal MS gate propagator, and the corresponding process error matrix $\tilde{\chi}_\mathrm{MS+}$ [Fig.~{\ref{fig:chi_MS+}}(c) and (d)] displays dominant error amplitudes in XX-II and II-XX elements, which are characteristic of an over-rotated propagator. We minimize the process error with a unitary propagator of form $U_\mathrm{MS+}=\mathrm{exp}(-i \theta^+ \mathrm{X}_1 \mathrm{X}_2)$ where $\theta^+ = 1.04(1)$~rad. This over-rotation produces an expected 6.2\% gate error, which is consistent with the decrease in the measured Bell state fidelity [$F_\mathrm{BST}^+ =  90.1(7)$\%] and process fidelity [$F_\mathrm{p}^+ =  81.6(5)$\%] compared to values for the optimized gate [$F_\mathrm{BST} =  96.2(7)$\%, $F_\mathrm{p}^+ =  88.1(5)$\%]. However, the process matrix provides significantly more information about the nature of the gate error that the Bell state tomography measurements do not.}

\section{Conclusion}
\label{sec:conclusion}

The greatest benefits to performing QPT are the sheer amount of information collected about a gate and the ability to identify prevalent error processes. The process matrix signature of a power miscalibration for an MS gate is demonstrated here, but other common gate errors such as unequal illumination of one ion over another or poor calibration of the red or blue axial sideband frequencies can also be diagnosed through process matrices. Two drawbacks of QPT are the extraneous errors due to the additional one-qubit rotations and its longer experimental duration in comparison to simpler evaluation techniques. Due to the longer duration, various experimental imperfections (e.g. laser noise, slow magnetic field drift) impact the process tomography protocol more than the Bell state tomography protocol, resulting in the fidelity estimate from process tomography being consistently lower than the Bell state fidelity \cite{gilchrist_distance_2005}. Bell-state tomography and QPT characterization methods are therefore complementary for diagnosing gate performance, and the adoption of one over the other depends on the specific needs of an experiment.

Quantum process tomography is a powerful technique for comprehensively characterizing quantum operations and diagnosing errors in their experimental implementation. \hnt{The potential modulation method demonstrated here for realizing single-ion rotations with a global gate beam is an effective method for performing QPT in trapped-ion systems; while the method can be extended to three or more ions with marginally larger potential changes and more complex pulse schemes \mbox{\cite{seck_single-ion_2020}}, it is particularly useful for studying architectures composed mostly of one-ion and two-ion quantum operations \mbox{\cite{Kielpinski_architecture_2002}}.} QPT analysis of identity and delay processes allows us to quantify the effects of laser noise and drift present in our experimental apparatus and to establish a baseline for use of the technique on more interesting two-qubit processes. We are able to reconstruct the process matrix of a two-ion entangling gate and to replicate the expected process errors for an intentionally miscalibrated gate. With this framework in place, it is possible to diagnose and mitigate specific errors in the MS gate, to predict the performance of longer gate sequences, and to study other quantum processes. \hnt{The low experimental overhead required for this technique will make QPT, along with alternative experimental protocols such as gate set tomography and randomized benchmarking, more accessible to other ion-trapping experiments.}

\ack

Research was sponsored by the Army Research Office under Grant Number W911NF-18-1-0166. The views and conclusions contained in this document are those of the authors and should not be interpreted as representing official policies, either expressed or implied, of the Army Research Office or the U.S. Government. The U.S. Government is authorized to reproduce and distribute reprints for government purposes notwithstanding any copyright notation herein. We would also like to thank Brian Sawyer for developing the fitting routines for analyzing two-ion fluorescence curves and Robin Blume-Kohout for discussions regarding MLE algorithms.

\section*{Appendix: Reduction of motional heating induced by potential modulations}
\label{sec:appendix}

To produce high-fidelity single-ion rotations for qubit state initialization and measurement during the QPT experiments, we sought to reduce the axial mode heating previously observed in randomized benchmarking experiments of the potential modulation single-ion addressing technique \cite{seck_single-ion_2020}. For that work, the trap electrode voltages were provided by National Instruments 16-bit PXI-6733 digital-to-analog converter (DAC) cards in a PXI-1045 chassis. The large number of cards in use there combined with an idiosyncrasy of the particular computer-chassis interface limited the update speed of the waveform voltages to 25 $\mu s$. To minimize the duration of the single-ion operations, voltages for the potential modulations were changed from initial to final values in a single update. This `snapped' potential heated the axial motional modes of the ions, and the heating could be reduced but not eliminated through calibration of background electric fields. Errors due to this motional excitation substantially increased for longer sequences of single-ion operations. 

\begin{figure}
    \centering
    \includegraphics[width=0.85\textwidth]{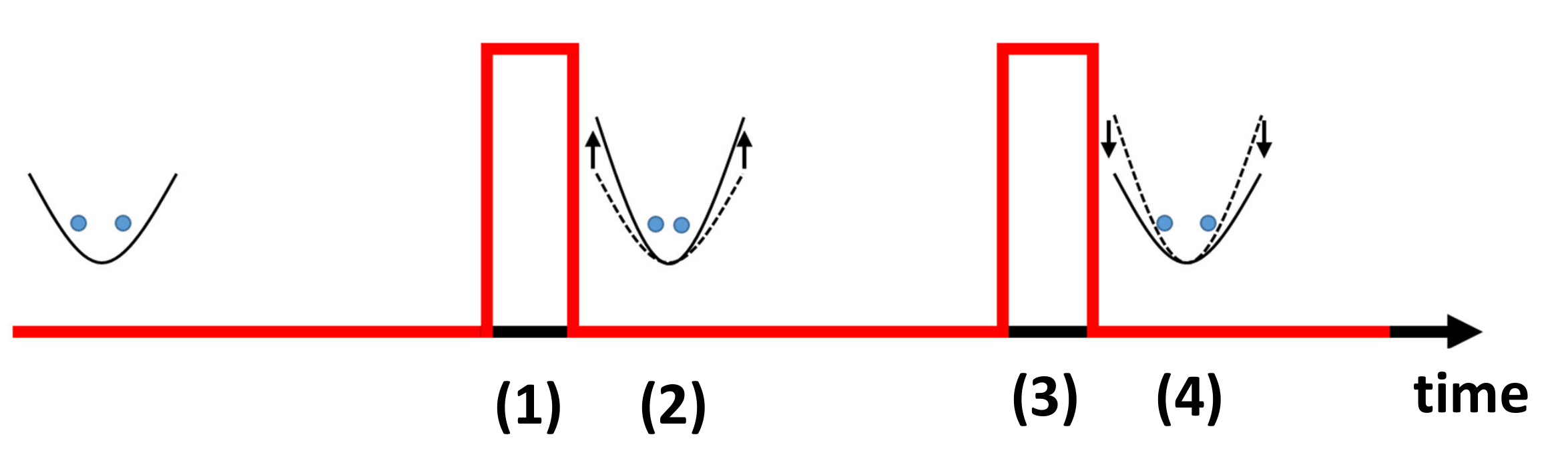}
    \caption {\label{fig:SIAPM}Pulse and potential-modulation sequence for the individual-ion-addressing method. Red lines indicate pulses of the 729 nm laser (steps 1 and 3) and black parabolas indicate configurations of and changes in the confining potential (steps 2 and 4). The interval from the initial laser pulse to the end of the second potential modulation (previously 50 $\mu s$ with the PXI-6733 cards) was reduced to \hnt{25 $\mu s$} with the PXIe-5413 cards.}
\end{figure}

To reduce this heating we installed DACs with a faster sampling rate to interpolate more smoothly between initial and final electrode potential values. We chose a system of National Instruments PXIe-5413 AWG cards (5 ns sampling rate) in a PXIe-1084 chassis to supply the trap voltages. This system enables the use of hundreds of interpolation steps between initial and final potential values within a small fraction of the duration previously required by the `snapped' potentials. The complete composite pulse sequence that produces one individually addressed single-ion rotation consists of two laser pulses and two potential modulations, as shown in Fig.~\ref{fig:SIAPM}. Beginning with the ions in an initial confining potential, (1) a single laser pulse ($\pi /2$ pulse for a total composite $\pi$ rotation or $\pi /4$ pulse for a total composite $\pi /2$ rotation) rotates the states of both ions, (2) the confinement is tightened with a change of electrode voltages thereby changing the ion positions and spacings, (3) a second laser pulse of the same duration as the first but with a different phase completes the rotation on the targeted ion and returns the untargeted ion to its original state, and (4) the confinement is returned to its original value to simplify preparation for the next arbitrary operation. 

\begin{figure}
    \centering
    \includegraphics[width=0.8\textwidth]{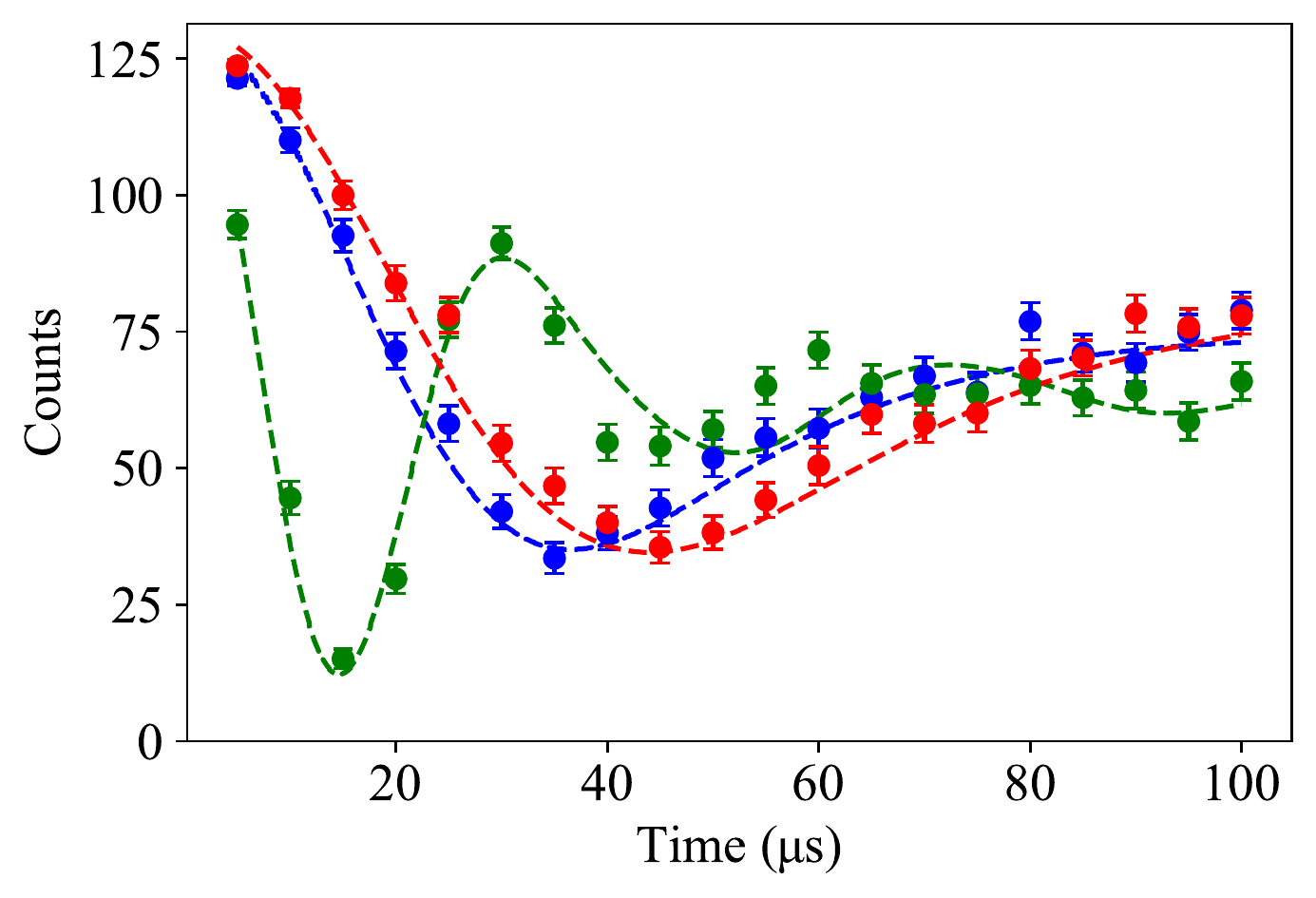}
    \caption {\label{fig:PM_heating}Rabi experiments on the COM blue motional sideband for a null waveform (blue), one cycle of a two-point potential modulation waveform (green), and 100 cycles of a 400-point potential modulation waveform (red). \hnt{Points represent measured two-ion fluorescence counts while dashed lines indicate fits to a heating model including thermal and coherent excitation contributions.}}
\end{figure}

To measure the heating caused by these potential modulations, we performed Rabi experiments on the axial COM and BM blue motional sidebands after allowing the ions to undergo some number of potential modulation cycles (departure and return to initial voltage values). The resulting two-ion fluorescence curves were fit to a temperature-dependent model to determine the occupation of the motional mode in coherent ($\bar{n}_\mathrm{coh}$) and thermal ($\bar{n}_\mathrm{th}$) distributions \cite{walther_controlling_2012, webster_thesis_2005}. Fig.~\ref{fig:PM_heating} shows such fluorescence curves for the COM mode for three different potential modulation cycles. The first `null' waveform (blue) involves no change of the trap confinement and serves as a baseline for the temperature of the mode (\hnt{$\bar{n}_\mathrm{th} = 5.5(4)$, $\bar{n}_\mathrm{coh} = 0.4(2))$}. The curve for the `snapped' two-point waveform (green) exhibits faster oscillations indicating a higher temperature (\hnt{$\bar{n}_\mathrm{th} = 32(4)$, $\bar{n}_\mathrm{coh} = 22(2)$}). For the QPT experiments presented above, we chose a waveform with 400 linearly interpolated samples and which exhibits no heating compared to the null waveform even after 100 potential modulation cycles (red, \hnt{$\bar{n}_\mathrm{th} = 3.5(3)$, $\bar{n}_\mathrm{coh} = 0.1(1)$}). Shorter waveforms were also investigated but were found to be less suitable, presumably because the 2 $\mu$s duration of the 400-point waveform is already comparable to the timescale set by the 530 kHz corner frequency of our trap electrode filters. We measured no heating of the breathing mode for either of these waveforms. These changes reduced the overall duration of single-ion composite rotations from more than 50 $\mu$s to \hnt{25 $\mu$s}, allowing us to produce faster sequences with higher fidelity for the QPT experimental framework.

\hnt{Incorporating transport into tomography protocols raises a concern about the effects of motional excitation on two-ion entangling gates or other processes under study. Phase gates are sensitive to the motional state of the ions and will experience errors of $\epsilon_\mathrm{th} = \frac{\pi^2}{4}\eta^4(\bar{n}_\mathrm{th}+2\bar{n}^2_\mathrm{th})$ for ions in a thermal state, where $\eta = \frac{1}{\sqrt{2}}\sqrt{\frac{\hbar}{2m\omega}}$ is the Lamb-Dicke parameter for the motional mode with axial frequency $\omega$ of two ions each with mass $m$ ($\eta = 0.039$ for the COM mode) \mbox{\cite{ballance_high-fidelity_2016}}. Our improved potential modulations reduce the motional excitation below a threshold that we can measure within our experimental error bars, even after 100 potential modulation cycles. If we bound any possible thermal excitation to the error bar of the fit (0.4 quanta), we can constrain the possible gate error contribution due to heating to $6\times 10^{-5}$. }

\clearpage

\end{document}